      [1999/12/01 v1.4c Il Nuovo Cimento]
\documentclass{cimento}

\title{Gravitino Dark Matter and the ILC}

\author{L.~Covi\from{DESY}}
\instlist{\inst{DESY} DESY Theory Group, Notkestrasse 85, D-22603 Hamburg, Germany}

\PACSes{\PACSit{04.60-m}{SUGRA}
\PACSit{12.60.Jv}{SUSY Models},
\PACSit{95.35+d}{Dark Matter}
}
\begin{document}

\maketitle

\begin{abstract}
We review the case of gravitino Dark Matter for stop, neutralino and sneutrino 
Next-to-Lightest Supersymmetric Particles and discuss prospects to 
investigate such scenarios at LHC and a Linear Collider.
\end{abstract}

\section{Introduction}

The gravitino, the superpartner of the graviton, is a well-motivated Dark Matter
candidate both from the theoretical and from the cosmological side: 
in fact it is naturally the Lightest Supersymmetric Particle (LSP) in many 
supersymmetry breaking mediation schemes, like gauge, gaugino or even 
gravity mediation, moreover a thermally produced gravitino in the 1-100 GeV 
mass range has the right Dark Matter (DM) energy density for a reheat 
temperature higher than that permitted if the gravitino is not the LSP and
decays via R-parity conserving couplings~\cite{gravitinodecay}. 
A gravitino DM and LSP scenario is nevertheless constrained, since
in this case the decay of the NLSP can endanger Big Bang Nucleosynthesis
(BBN). The lifetime of the NLSP, for conserved R-parity, is fixed by the
supersymmetry breaking masses without any free parameter and 
for a gaugino neutralino or stop NLSP (for the higgsino case it can be 
approximately a factor 2 longer if the heavy Higgs decay channels are 
kinematically closed) is given by
\begin{equation}
\tau_{NLSP} = 60 s \left( \frac{m_{NLSP}}{1\;\mbox{TeV}} \right)^{-5} 
\left( \frac{m_{3/2}}{10\;\mbox{GeV}} \right)^2 
\end{equation}
So we see immediately that for gravitino masses of 10-100 GeV and 
NLSP masses within the range of the LHC and ILC, the decay happens during 
or after Nucleosynthesis and care has to be taken to avoid  destroying the 
successful BBN predictions.

There are though two natural ways to relax these contraints apart from
invoking R-parity breaking as in~\cite{bchiy07}:
either the NLSP is strongly interacting and its number density is therefore
reduced in comparison to a weakly interacting massive particle (WIMP), or,  
if the NLSP is a WIMP, like the neutralino or sneutrino, its decay channels 
or the spectra of the other superpartners are appropriately chosen to make it 
as harmless as possible.
In these proceedings we will shortly review the situation for a stop \cite{bckp08},
neutralino~\cite{chpr09} and sneutrino~\cite{ck07}  NLSPs and discuss possible 
signatures of these NLSPs at colliders.


\section{BBN constraints on decaying particles}

The bounds on the decay of a heavy neutral particle during or after BBN have
been the object of careful study~\cite{BBN-bounds}.
It was found that if a particle releases a too large energy in either hadronic 
or electromagnetic showers, it can destroy the more fragile light nuclei like 
Deuterium and Lithium or split the much more abundant Helium nuclei and 
increase other species.
Since Nucleosynthesis proceeds in steps starting from approximately 0.1~s
after the Big Bang, these constraints depend not only on the NLSP energy 
density, but also on its exact time of decay.
For a particle of 1~TeV mass, the hadronic showers bounds are approximately 
given by~\cite{had-BBN}
\begin{eqnarray}
 \Omega_{NLSP} h^2 \leq \left\{
\begin{array}{cl}
2.3\times 10^{-1} & \mbox{for}\quad  0.1 \; \mbox{s}\leq  \tau_{NLSP} \leq 10^2\; \mbox{s} \cr
2.8\times 10^{-5}  & \mbox{for}\quad 10^2\; \mbox{s} \leq \tau_{NLSP} \leq 10^7\; \mbox{s} \cr
2.8\times 10^{-6}  & \mbox{for}\quad  10^7\; \mbox{s} \leq \tau_{NLSP} \leq 10^{12}\; \mbox{s} \cr
\end{array} \right.
\nonumber
\end{eqnarray}
where $ \Omega_{NLSP} h^2 $ is the energy density  that the NLSP would have today
if it had not decayed in units of the critical density, and we assume a branching ratio into 
hadrons of order one.
The electromagnetic bounds are less severe for short lifetimes, but become comparable
for $  \tau_{NLSP} \geq 10^6 $ s.
Note that if the particle is electromagnetically charged and has a low hadronic
branching ratio, like the stau, also constraints from catalyzed BBN \cite{BBN-bounds} 
are important, but for the stop they are weaker than the bounds above.
On the other hand recently very stringent BBN constraints have been derived 
for strongly interacting neutral relics that couple to the nuclei like a nucleon and 
can form nuclear bound states~\cite{sBBN}. These constraints apply to a stop NLSP
and are stronger than the hadronic shower ones given above for lifetimes longer 
than $ ~ 200 $ s. They exclude strongly interacting relics for densities above 
$ \Omega_{NLSP} h^2 \sim 10^{-6}-10^{-7} $ from $ \tau_{NLSP} \geq 300 $ s,
 reaching down even to $ \Omega_{NLSP} h^2 \sim 10^{-10}-10^{-11} $ from 
$ \tau_{NLSP} \geq 2\times 10^3 $ s.

\section{Stop NLSP}

The lighter of the two superpartners of the top quark can be one of the lightest
particles of the SUSY spectrum thanks to the large left-right mixing in the
soft supersymmetry breaking scalar mass matrix. At the same time it is a 
strongly interacting particle and it may therefore be expected to freeze out 
with a sufficiently small number density before decaying into the gravitino.

We computed the number density of such a scalar particle concentrating on the
two gluons final state in \cite{bckp08}; this channel has the advantage of being 
independent of the supersymmetric parameters apart for the stop mass and of 
being enhanced by the Sommerfeld effect.  We found indeed that this enhancement 
factor is not negligible and can increase the annihilation cross-section into gluons 
by a factor of order 2-3;  therefore usually the gluon channel is the 
dominant contribution to stop annihilation cross-section. 
Nevertheless the annihilation process at freeze-out is not efficient enough to 
relax all the BBN bounds:  it reduces the number densities so that a stop NLSP 
is consistent with BBN for masses below 1 TeV and lifetimes less than 100 s
(corresponding to a maximal gravitino mass of  the order of 10 GeV for
a 1 TeV stop mass).

For longer stop lifetimes the bound on its number density is stronger and 
needs additional annihilation to take place at the QCD phase transition.
If the annihilation there reaches the level of the unitarity cross-section, 
increasing to  $\sigma \propto 1/m_{\tilde t}^2 $ due to $ \alpha_{QCD} \sim 1 $  
at  temperatures $ T \sim \Lambda_{QCD} $, then the stops  can return in thermal 
 equilibrium and start annihilating again, 
as long as $m_{\tilde t} $  is approximately below 700 GeV. 
In fact the condition for the stops to reenter thermal equilibrium before or
at  $ T \sim \Lambda_{QCD} $ is given approximately by
\begin{equation}
n_{\tilde t} \langle \sigma v \rangle_{T=\Lambda_{QCD}} \geq H(\Lambda_{QCD})
\quad\Rightarrow \quad
m_{\tilde t} \leq \frac{\Lambda_{QCD}}{ x_f \alpha_{QCD}^2( x_f)}
\sim 700\; \mbox{GeV}\;,
\end{equation}
where $ x_f = T_f/m_{\tilde t} \sim 1/35 $ is the freeze-out temperature for the first 
freeze-out process and $ \alpha_{QCD}(x_f) \sim 0.1 $ is the QCD coupling constant 
at that temperature.
If this condition is satisfied, the stop number density is strongly reduced
and reaches down to
$ \Omega_{NLSP} h^2 \sim 3 \times 10^{-6} \left( m_{\tilde t}/700 \mbox{GeV} \right)^2 $
This value is below the BBN hadronic shower bounds  for lifetimes up to $ 10^7 s $,
and would seem therefore to allow for any gravitino mass below the stop mass.
Unfortunately though, this is not sufficient to evade the more recent strongly interacting 
relic constraints~\cite{sBBN}. 
For small stop masses though, a small windows remains open up to lifetimes of the 
order $~ 2-300 $ s, i.e. a gravitino mass of 1-10 GeV for  $ m_{\tilde t} \sim 250-700 $ GeV.
Note that this conclusion does not change qualitatively even if the stop annihilation 
rate at the QCD phase transition becomes of the order of the pion cross-section
as invoked in \cite{Luty08}, since then the stop energy density is reduced by 
another order of magnitude, but still in conflict with the strongly interacting relic 
BBN constraints as discussed in~\cite{sBBN}.

We have therefore two BBN consistent scenarios, one with a stop NLSP
below 1 TeV with a light gravitino $ m_{3/2} \leq 10 $ GeV, while the other with
a very heavy stop such that is decays in any case before BBN.  
In the first case the stop NLSP is within reach of the LHC and, being very
long-lived on collider timescales,  should hadronise and be seen there as a 
strange hadron or meson~\cite{stop-hadron}.
Unfortunately for this scenario, a 500 GeV linear collider will not be able to 
produce any supersymmetric particle since they would be outside its energy 
range. In fact the CDF collaboration has already performed searches for
such particles at Tevatron and excluded a metastable stop with mass
below 249 GeV~\cite{stop-CDF}.

\section{General neutralino NLSP}

A neutralino NLSP is a WIMP and has usually a quite large number density 
at freeze-out. In fact, within the constrained MSSM, where the lightest state is
mostly bino, it is excluded by BBN for gravitino  masses above 1 GeV or 
so~\cite{CMSSMneutralino}.
We investigated the BBN bounds on a general neutralino, varying its mass and 
composition between bino, wino and higgsino in \cite{chpr09}.
We varied the gaugino mass parameters $ M_{1,2} $ and the higgsino mass
parameter $ \mu $ independently and scanned so over different compositions
and masses for the neutralino. We imposed the LEP constraints on the
chargino mass and required the neutralino to be always lighter than the
charginos. We used the MICROMEGAS package \cite{micro} to compute
the number density of the neutralino NLSP.

We explored two strategies to relax the BBN bounds: either tuning the
neutralino composition to give the minimal branching ratio into hadrons or 
enhancing the annihilation cross-section to reduce the yield at freeze-out 
as much as possible.
The first strategy is unfortunately not successful since the neutralino
with the lowest hadronic branching ratio, a pure photino, has in any case 
a too large number density at freeze-out.
On the other hand the second strategy pointed to the case of a mainly 
higgsino or purely wino NLSP, for which the number density can be quite 
strongly reduced.
In fact for a light wino, just above the LEP chargino bound and nearly 
degenerate and therefore coannihilating with the charginos,  
the freeze-out density and the hadronic branching ratio are both reduced
so that gravitino masses up to 5-10 GeV become allowed.
Another interesting parameter region is the place where the annihilation 
cross-section is resonantly enhanced due to the heavy Higgs s-channel. 
In this case, which requires a substantial higgsino component and
$ 2 m_\chi \sim m_{A,H} $, the bounds can be bypassed for lifetimes 
shorter than $10^2$ s. So a gravitino mass  e.g. up to 70 GeV is allowed 
for neutralino masses around 1 TeV. 

Both these new allowed regions, nearly degenerate light wino and 
higgsino annihilating at the Higgs resonance, undoubtly require the 
precision of a linear collider in the mass determination either to disentangle 
the different mass eigenstates or to distinguish this scenario from the 
case of neutralino LSP and DM annihilating also resonantly, but at the
fringes instead than on top of the resonance.

\section{Sneutrino NLSP}

A sneutrino NLSP has the great advantage with respect to other sparticles
that it decays mostly in gravitino and neutrino and neutrinos are relatively
harmless for BBN since they cannot interact electromagnetically or hadronically.
In general therefore the BBN bounds on the sneutrino number densities
are much weaker than for other MSSM superpartners~\cite{BBN-snu}.

It is usually not easy to obtain a sneutrino NLSP since if one assumes
universality at the GUT scale,  the RGE running tends to make all left-handed (LH)
electroweak multiplets degenerate, apart for small D-term mass contribution, 
and the right-handed states lighter than the LH. 
Therefore to have a sneutrino NLSP one has to relax universality. 
For example one can choose non universal Higgs masses, 
as it happens in case of gaugino mediation in extradimensional models~\cite{bcks06}, 
and obtain smaller masses for the LH neutral states, but nearly degenerate with 
the charged LH sfermions and also the neutralino.
We studied the region of sneutrino NLSP in this scenario \cite{ck07} and
found out that coannilation plays an important role and increases the
sneutrino number density in some regions so that BBN constraints can become 
relevant even for small NLSP masses. 
On the other hand a large parameter region is consistent with BBN and has light 
sneutrinos and very small mass splitting between the charged sleptons of all three  
generations and the lightest neutralino. Due to the very degenerate spectra, 
many of the light SUSY particles can decay only via off-shell W and 3-body 
final states~\cite{snu3body}, characterized by a large number of leptons, also
of different flavour, in the final state. This signature seems to be promising for
a search of sneutrino NLSP at the LHC \cite{snu-LHC}.
On the other hand, such very near-degenerate (neutral)  states will be difficult to 
disentangle at the LHC, so that also in this scenario the Linear Collider could 
provide precise mass measurement and allow possibly to distinguish 
between sneutrino NLSP and neutralino LSP.

\section{Conclusions}

Gravitino Dark Matter is a well-motivated scenario, which suffers from
quite strong constraints from BBN as soon as the NLSP lifetime becomes
longer than 1 s, i.e. for 100 GeV neutralino masses, for gravitino masses
larger than 1 GeV or so.
We have discussed here a couple of NLSP scenarios that can satisfy all
the constraints with larger gravitino masses: 
a relatively light stop NLSP, a light wino neutralino NLSP or a neutralino 
with large higgsino component which can resonantly annihilate via 
heavy Higgs exchange or a LH sneutrino NLSP.
In the last cases the allowed regions are characterised by very small 
mass differences either between super-partners of the EW multiplets or
between the neutralino and half the heavy Higgs masses.
In such cases the linear collider precision in mass determination
may become vital to disentangle such scenarios, even if some initial 
supersymmetric signature should (hopefully !) be seen at the LHC.

\acknowledgments

The author would like to thanks C.~F.~ Berger,  J.~Hasenkamp,
S.~Kraml, F.~Palorini, S.~Pokorski and J.~Roberts for the very fruitful 
and enjoyable collaborations.
The author would also like to thank the organisers and INFN for 
the financial support and for  the stimulating atmosphere during 
the workshop.

This work has been supported by the "Impuls- und Vernetzungsfond" 
of the Helmholtz Association under the contract number VH-NG-006.

\end{document}